\documentclass[aps,prl,reprint,groupedaddress]{revtex4-2}
\usepackage{graphicx}
\usepackage{dcolumn}
\usepackage{bm,epstopdf,soul}
\usepackage{amssymb,amsfonts,amsmath}

\begin{document}

\title{Spin memory of the topological material under strong disorder}
\author{Inna Korzhovska,$^1$ Haiming Deng,$^1$ Lukas Zhao,$^{1}$ Zhiyi Chen,$^1$ Marcin Konczykowski,$^2$ Shihua Zhao,$^1$ Simone Raoux,$^3$ \& Lia Krusin-Elbaum$^{1,\dagger}$}
\vspace{10mm}
\affiliation{$^1$Department of Physics, The City College of New York - CUNY, New York, NY 10031, USA}
\affiliation{$^2$Laboratoire des Solides Irradi\'{e}s, \'{E}cole Polytechnique, CNRS, CEA, Universit\'{e} Paris-Saclay, 91128 Palaiseau cedex, France}
\affiliation{$^3$Helmholtz-Zentrum Berlin f\"{u}r Materialien und Energie, 12489 Berlin, Germany}

\date{\today}

\begin{abstract}
	Robustness to disorder -- the defining property of any topological state -- has been mostly tested in low-disorder translationally-invariant materials systems where the protecting underlying symmetry, such as time reversal, is preserved. The ultimate disorder limits to topological protection are still unknown, however a number of theories predict that even in the amorphous state a quantized conductance might yet reemerge. Here we report a directly detected robust spin response in structurally disordered thin films  of the topological material Sb$_2$Te$_3$  \emph{free of extrinsic magnetic dopants},  which we controllably tune from a strong (amorphous) to  a weak (crystalline) disorder state. The magnetic signal onsets at a surprisingly high temperature ($\sim 200~\textrm{K}$) and eventually ceases within the crystalline state.
	We demonstrate that in a strongly disordered state
	\emph{disorder-induced spin correlations} dominate the transport of charge --- they engender a spin memory phenomenon, generated by the nonequilibrium charge currents controlled by localized spins.
	The negative magnetoresistance (MR) in the extensive spin-memory phase space is isotropic.  Within the crystalline state, it transitions into  a positive MR corresponding to the weak antilocalization  (WAL) quantum interference  effect, with a 2D scaling characteristic of the topological state.
	Our findings demonstrate that these nonequilibrium currents  set a disorder threshold to the topological state; they lay out a path to tunable spin-dependent charge transport
	and point to new possibilities of spin control by disorder engineering of topological materials
\end{abstract}

\maketitle


Electronic disorder \cite{Anderson1979} and elementary excitations in quantum condensed matter are fundamentally linked 
and it is well established that spatially fluctuating potentials tend to promote decoherence and localization of fermions, i.e. formation of Anderson insulators \cite{Mirlin2008}.  
The interplay of interactions and disorder often leads to new quantum behaviors; disorder typically boosts interparticle correlations both in charge and in spin channels, and that could either aid or suppress the motion of charge \cite{Mott1997}.
Spin effects related to disorder are particularly important when  spin-orbit coupling (SOC)  is strong \cite{Winkler2003}, and when spin-dependent charge transport can be electrically manipulated for uses, e.g. in spin-based electronics \cite{DasSarma2004}.

Strong SOC is a hallmark of three-dimensional (3D) topological insulators \cite{Qi2011},  where
2D gapless spin-polarized Dirac surface states are robust against backscattering.
Most topological materials are known to contain a natural population of charged defects \cite{Scanlon-antisites2012} that do not cause a destruction of the topological Dirac states \cite{Beidenkopf2011}; indeed, they can be compensated
\cite{irrad-Lukas2016} as long as Dirac mass \cite{MagDisorderDavis2015} or puddle \cite{Ando-negMR2017} disorders do not enter. Under weak disorder, a coherent interference of electron waves survives disorder averaging \cite{Bergmann1984}, and  strong SOC enhances conductivity by a weak antilocalization (WAL) correction related to the topological $\pi$-Berry phase \cite{Qi2011} when magnetic impurities are absent. The 2D WAL channels can be  outnumbered by the weak localization (WL) channels  \cite{KapitulnikWAL2013} --- this is a precursor of Anderson localization \cite{Mirlin2008}, which occurs at strong disorder.
Under strong disorder, theory and numerical simulations \cite{Li2009,Beenakker2009,Franz2010,Vojta2012} predict an emergence of a new topological state, dubbed `topological Anderson insulator', in which conductance $G_0 = 2e^2/h$ is quantized. Indeed, recent theoretical demonstrations of topological phases in amorphous systems \cite{amorphTI-NP2018,amorphTI-PRL2017} point to promising new possibilities in engineered random landscapes. Strong disorder, however, is not trivial to install, quantify and control,  and topological matter under such conditions has not yet been experimentally tested.

Here we implement an extensive range of site disorder -- from amorphous to crystalline state -- in Sb$_2$Te$_3$, the material which
is a known 2$^{nd}$ generation topological insulator \cite{Zhang-NatPhys09,ARPES3,SpinCusp-Lukas2014}, and report that under strong structural and electronic disorder conditions dynamic spin correlations dominate charge transport over a surprisingly large range of magnetic fields.
These correlations imprint spin memory on the {electrons hopping 
via localized spin sites}. Predicted to be small \cite{Aleiner-Spivak2014} and thus practically unobservable in the conventional materials, the effect found here is large; it persists over a surprisingly wide range of disorder and well within the disordered crystalline topological phase,
as long as variable range electron hopping (VRH) \cite{Shklovskii-Efros1984} is at play.  It  eventually transitions into the characteristic 2D WAL regime when the gapless surface channels are reestablished.  The spin memory uncovered in this work is \emph{not an orbital effect}; it is distinct from the weak localization interference effect (WL)  observed in the magnetically doped topological insulators \cite{WL-WAL-Xue2012}.  It originates from the presence of disorder-induced localized spins \cite{Mott1997} and, as witnessed by the characteristically non-analytic negative magnetoresistance (neg-MR), is governed by the distribution of very large spin $g$-factors that widens with decreasing localization length $\xi$ in a way akin to an assembly of quantum wells \cite{Chen2007}.

The experiments were performed on thin ($\sim 20-50$ {nm}) films of Sb$_2$Te$_3$, in which extreme positional disorder (amorphous state) is possible to obtain (SI, Section \textbf{A}). Sb$_2$Te$_3$ is a well known phase-change material \cite{LiaPCM2009} {(SI, {Fig.~S1)}, that undergoes amorphous-to-crystalline transformation with the concurrent orders-of-magnitude resistive drop, and hence a huge range of disorder could be controllably explored.
	
	\begin{figure*}
		\centering
		\includegraphics[width=\textwidth]{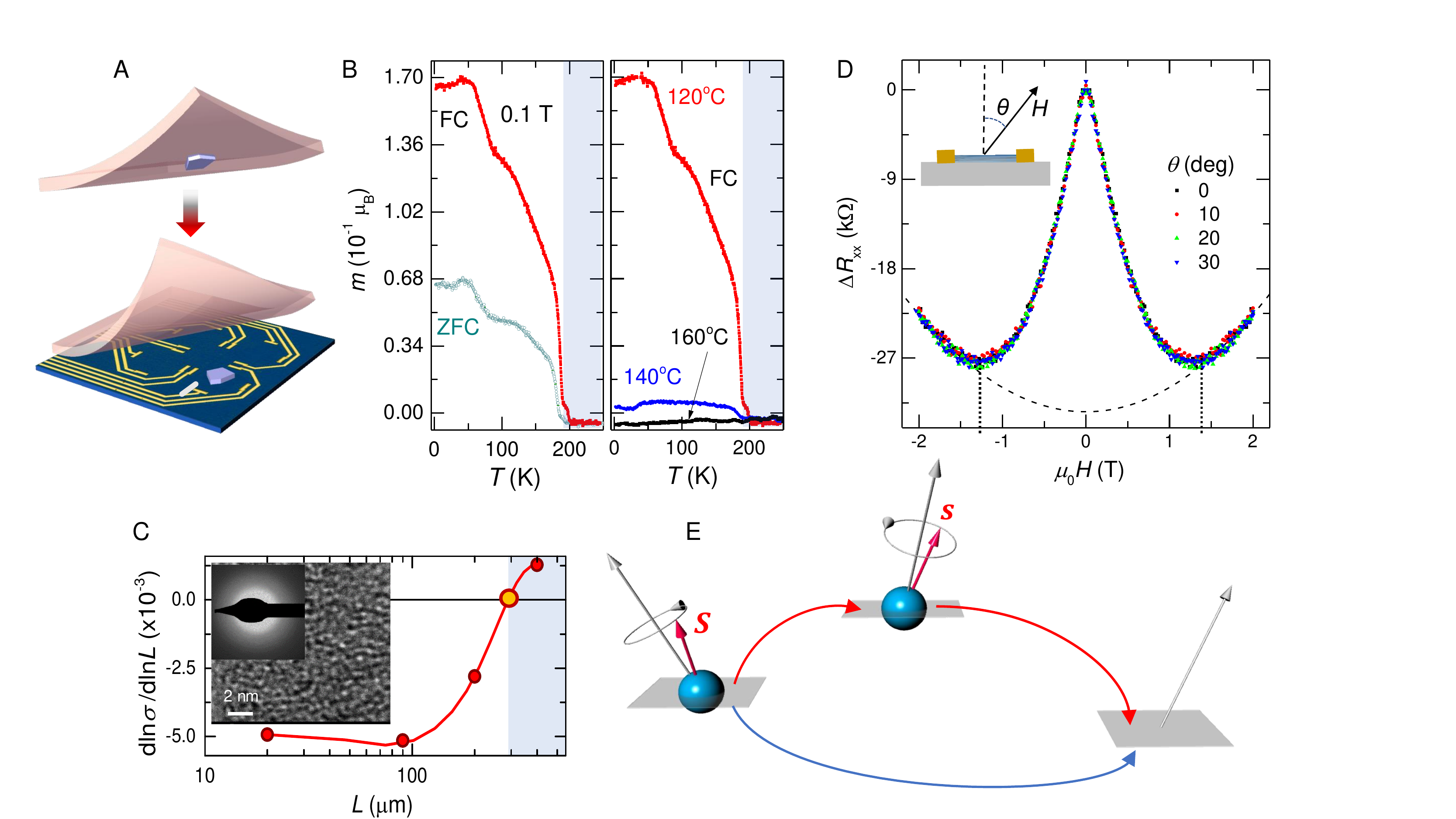}
		\caption{{Non-equilibrium spin correlations in strongly disordered Sb$_2$Te$_3$.}
			{(A)}, Thin (20 nm) Sb$_2$Te$_3$ flakes were exfoliated and transferred atop a $\mu$-Hall sensor (Methods). The scale bar is $50~\mu\textrm{m}$.
			{(B)}, Magnetic moment per atom, $m$, measured by a $\mu$-Hall sensor. \textit{Left:} $m$ in a 0.1 T field under field-cooling (FC) and zero-field-cooling (ZFC) conditions. \textit{Right:} $m$ of the same flake for three levels of disorder indexed by annealing temperature $T_{a}$. It onsets at $\sim 200$ K, saturates at low $T$, and becomes vanishingly small for  $T_{a}\gtrsim 160^\circ$C .  {(C)}, As-deposited Sb$_2$Te$_3$ films are amorphous as seen by the transmission electron microscopy (inset). The lateral size $L$ dependence of longitudinal conductivity $\sigma$ obeys Anderson scaling \cite{Anderson1979}.
			{(D)}, Longitudinal resistance $R_{xx}(H)$ of
			strongly disordered Sb$_2$Te$_3$ films ($T _a = 120^\circ$C)  does not depend on the tilt of magnetic field (see sketch). The dissipation reduction is relatively large (several $\%$); it is observed below $H_{max}$, the field at which $\Delta R_{xx}(H)$ crosses over to the $H^2$-dependence (dash).
			Data in ({C}) and ({D}) were taken at 1.9 K.
			{(E)}, Cartoon of electron hopping to a distant empty site under flowing current, see text. The electron with spin $\textbf{\textit{S}}$ may reach this site directly or via an intermediate site hosting localized spin $\textbf{\textit{s}}$. Random disorder landscape induces spatial randomness in the spin $g$-factors and hence the randomness of local fields (grey arrows) controlling the precession of all spins. The indirect channel will depend on the state of $\textbf{\textit{s}}$ during hopping attempts and the memory of this state is encoded in the transport of charge. These spin correlations are destroyed by magnetic field, and consequently the resistance is reduced.
		}
		\label{fig1}
	\end{figure*}
	
	Let us recall that in the presence of disorder a finite population of singly occupied states below the Fermi energy $E_F$ has been discussed as long as 20 years back by Sir Neville Mott \cite{Mott1997}.
	Magnetic response from randomly localized spins in such state was expected to be weak and, as far as we know, has never been experimentally demonstrated. So our first surprising finding was a very robust magnetic signal from the disordered Sb$_2$Te$_3$ films directly detected using a custom-designed $\mu$Hall sensors (Figs.~1{A},{B}), with the thin film flakes mechanically exfoliated from their substrates and transferred onto the active sensor area ({Materials and Methods, SI, Fig.~S2}). Our films do obey Anderson scaling (see Fig. 1{C}), and we surmise that here the observed effective moment per atom is significantly amplified by the large effective Land\'{e} \textit{g}-factor \cite{Winkler2003} in Sb$_2$Te$_3$, where SOC is strong \cite{Zhang-NatPhys09}. The detected signal depends on the magnetic history (field-cooling \textit{vs.} zero-field cooling), which, together with slow magnetic relaxation (Fig.~ S2C) reflects a glassy nature of the localized state. We emphasize that magnetic signal crucially depends on the level of disorder; indeed, it becomes barely detectable in the crystalline phase as disorder in \emph{the same film} is reduced by thermal annealing (Fig.~1{B}).

	\begin{figure*}[tb]
		\centering
		\includegraphics[width=\textwidth]{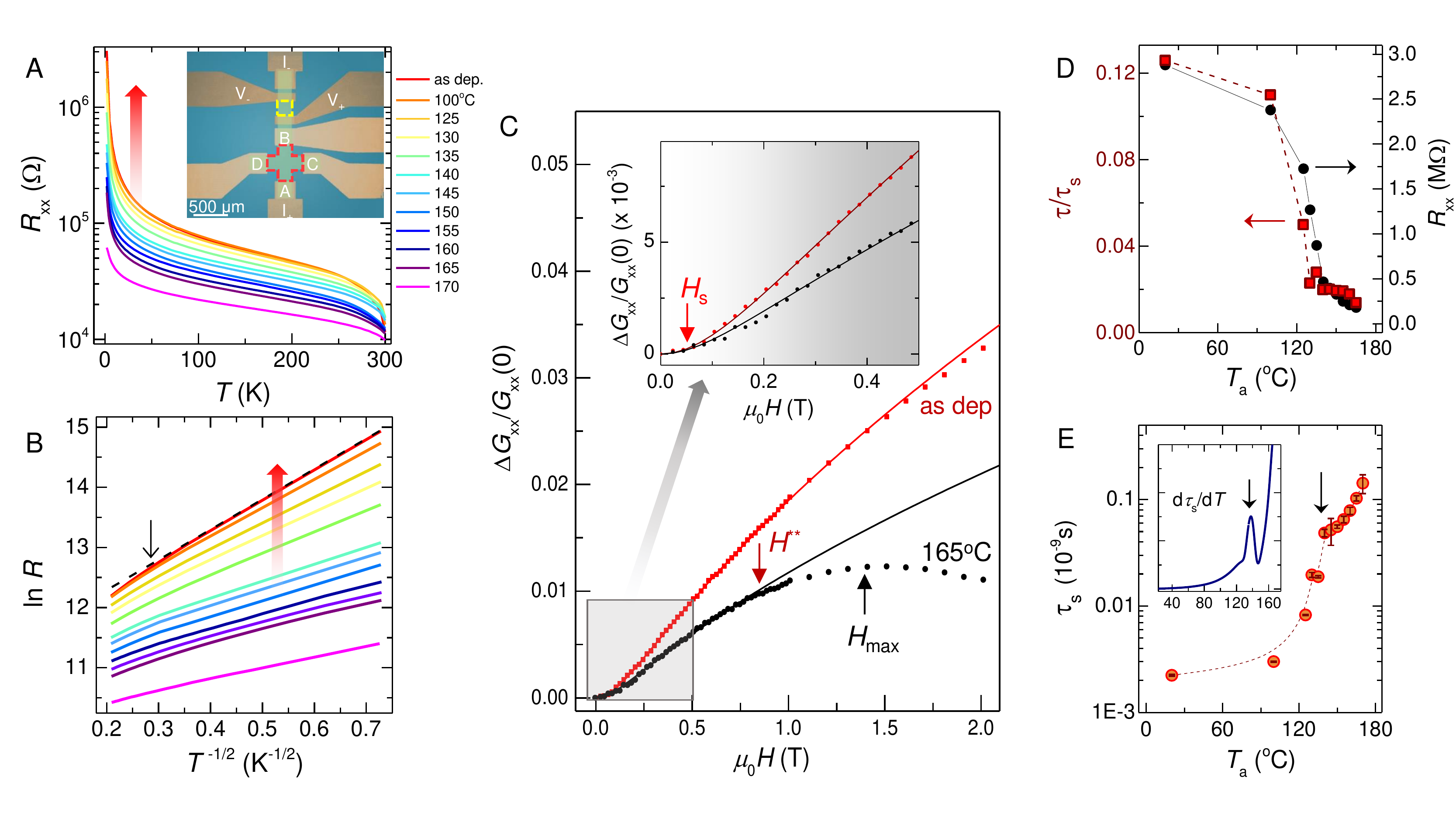}
		\caption{{Spin memory effect in the longitudinal conductance $G_{xx}$ of Sb$_2$Te$_3$ in the VRH regime.}  {(A)},   $R_{xx} (T)$ of a 20~nm thin Sb$_2$Te$_3$ film under strong disorder increases at low temperatures by orders of magnitude; the strength of disorder is reproducibly controlled by $T_a$. Inset: lithographically patterned  Hall-bar and van der Pauw contact configurations used in measurements of $R_{xx}$. {(B)},  $R_{xx}$ exponential in inverse $\sqrt{T}$ confirms that below $\sim 10$~K charge transport is by the 3D VRH \cite{Shklovskii-Efros1984}, the regime where dynamic spin correlation are expected. The color code is as in {(A)}). Red arrows in ({(A)} and {(B)}  point in the direction of increasing disorder.
			{(C)}, Field dependence of $G_{xx}$ for two disorder states shown at 1.9 K. A fit (solid lines) to Eq. (1) fully reproduces the non-analytic form of $\Delta G_{xx}(H)/G_{xx}(0)$ arising from spin-memory in the VRH regime. Inset: Zoom of the data at low fields.
			{(D)}, Disorder dependence of $\tau/\tau_s$, the ratio of hopping to spin-relaxation time, is found to closely correlate with the low-$T$ $R_{xx}$.
			{(E)}, $\tau_s$ increases by nearly two orders of magnitude in the disorder range studied; it is a very sensitive probe of the phase change at crystallization, see inset.
		}
		\label{fig2}
	\end{figure*}
	
	\begin{figure}[tb]
		\centering
		\includegraphics[width=\linewidth]{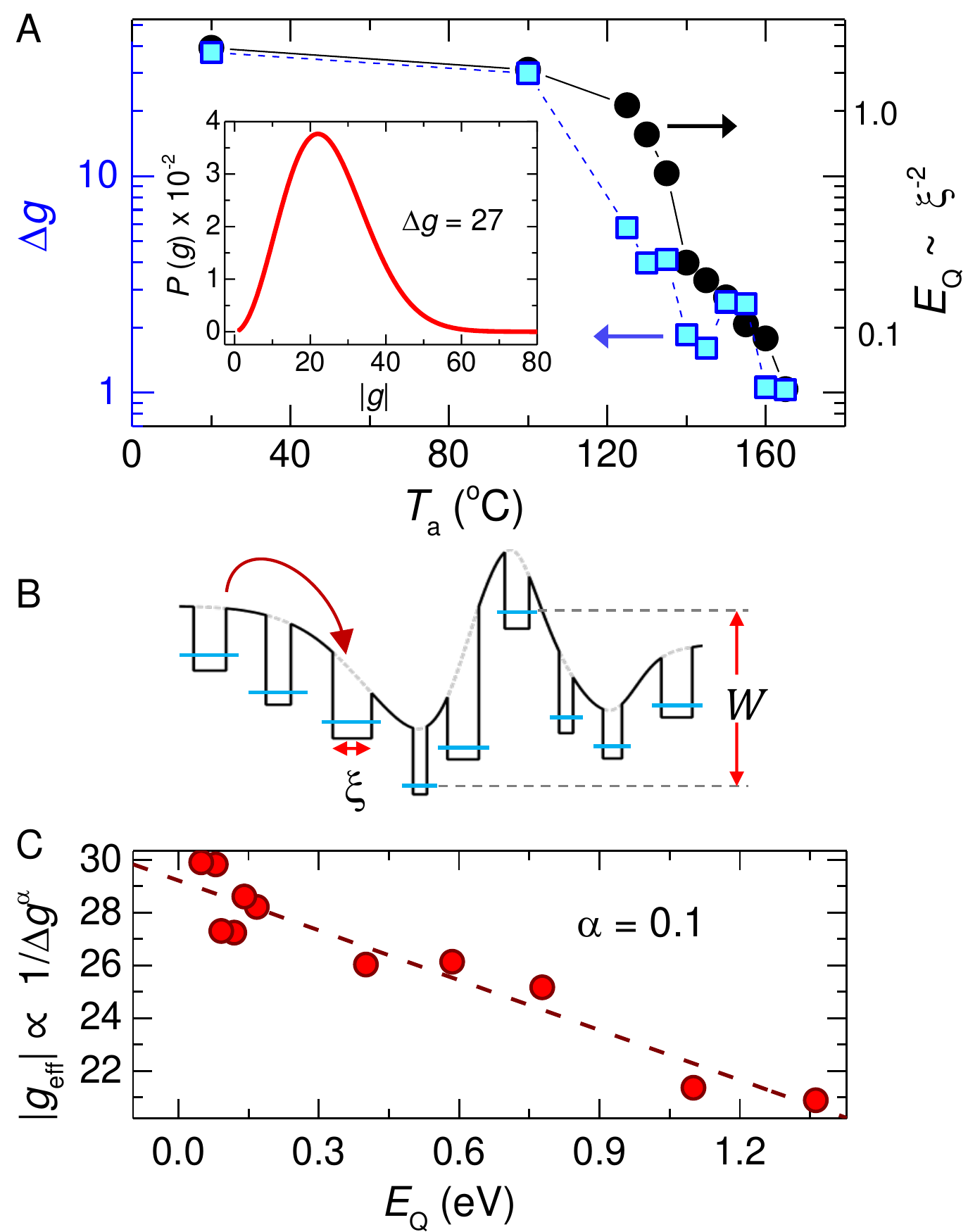}
		\caption{ {Distribution of spin $g$-factors and confinement energy.}
			{(A)},  Width $\Delta g$ of the $g$-factor distribution $P(g)$  \textit{vs}. $T_{a}$
			follows the confinement energy $E_Q(T_{a})$. $E_Q$ is modeled by a particle-in-a-box and calculated using $\xi$
			as a box size. In the amorphous state in the conduction tail states effective mass is strongly enhanced \cite{eff-mass-amorph2002}. Inset:  $P(g)$ calculated in the strong SOC regime \cite{Csonka2008} for $\Delta g = 27$.  {(B)},  A model of
			disorder landscape within the  Anderson bandwidth $W$ riding on a long-range smooth potential \cite{Gantmakher2005}.
			{(C)},  Effective $g$-factor, $|g_{eff}|$, increases  with $E_Q$. We note that the sign of $g$ for many semiconductors, particularly with strong SOC, is negative \cite{Winkler2003}.
		} \label{fig3}
	\end{figure}
	
	Our second key finding is shown in Fig. 1{D}. The change in the longitudinal magnetoresistance $\Delta R_{xx}$ at low magnetic fields is with a remarkable fidelity impervious to the tilt of magnetic field. It is relatively large and negative (i.e. charge transport becomes less dissipative) up to a field $H_{max}$ (Figs.~1{D} and ~S3). This field isotropy of the low-field dissipation `quench' naturally suggests a spin-dominated mechanism rather than orbital effect, and so the pertinent question to ask is how such behavior can proceed in a system where electronic states are localized and the \emph{extrinsic} spinfull impurities are absent ({Table S1}).
	
	With strong disorder charge transport is a complex electron hopping process that at low temperatures proceeds via quantum tunneling between localized states assisted by phonons \cite{Shklovskii-Efros1984,Gantmakher2005}. While considerations of magnetotransport have mainly focused on the orbital effects, a recently proposed idea \cite{Aleiner-Spivak2014} takes note of putative nonequilibrium spin correlations in the localized regime created by the flowing current when electron hopping times $\tau$ are short relative to the spin relaxation times $\tau_s$. These time scales determine the magnetic field range over which spin-correlation-driven neg-MR (positive magnetoconductance) ought to be present. The idea is illustrated in Fig. 1{E}. When the current is injected, an electron with spin $\textbf{\textit{S}}$ attempting to hop to an available empty site can do it in two ways: directly or via an intermediate site occupied by a localized spin $\textbf{\textit{s}}$. It may take several attempts for the indirect hops to succeed and the return probability will depend on history, i.e. on whether the tunneling electron can form a triplet or a singlet state with $\textbf{\textit{s}}$.  For example, in the absence of disorder a triplet state would remain so in the presence of applied magnetic field and no reduction of magnetoresistance (increase of magnetoconductance) is expected. Under strong disorder (such as shown in Fig.~1{C}), however, spin $g$-factors will be spatially random so that localized spins at different sites will precess incoherently and spin correlations will be destroyed by the field.  Accordingly, in a simple model \cite{Aleiner-Spivak2014} the change in magnetoconductance $\Delta G_{xx}$ arising from such spin correlations should follow not a power law \cite{Joffe-Spivak2013} but a unique non-analytic form:
	\begin{equation}
	\begin{split}
	\frac{\Delta G_{xx}(H)}{G_{xx}(0)} 
	&\sim A \Bigl[-\Gamma\left(-\frac{d_s}{2}\right)\Bigr] \sum_{l = -1}^{1} \Bigl[{ \left({ \frac {i l H}{H^{\star\star}} + \frac{\tau}{\tau_s}} \right)^{d_s/2}} \\
	&-{\left( \frac{\tau}{\tau_s}\right)}^{d_s/2}\Bigr],
	\end{split}
	\end{equation}
	where $A = \frac {G_{xx}(H\rightarrow \infty) - G_{xx}(0)}{G_{xx}(0)}$, $d_s  =  4/3$ is the spectral dimension of the percolation cluster \cite{percolation2002} (which is the relevant dimension in the hopping process), $l$ is  index of the diffusing spin, $\Gamma(-\frac{d_s}{2}$) is the gamma function $ \cong - 4$,
	and $H^{\star\star} = \frac {\hbar}{\mu_B\tau\Delta g}$  is the limiting magnetic field range set by the hopping rate $1/\tau$ and the disorder-induced spread $\Delta g$ of spin $g$-factors.
	
	The strongly localizing behavior we observe in the longitudinal resistance $R_{xx}$ (Fig. 2{A}) at low temperatures (below $\sim$ 10 K) follows variable range hopping law $ R_{xx} (T)  = R_0 \textrm{exp} {({\frac{T_0}{T}})}^{1/2}$ of Efros-Shklovskii (E-S) kind \cite{Gantmakher2005}, see Fig.~2{B}. The E-S energy scale $T_0$ characteristic of the hopping process {(Fig. S1B)} is tracked on decreasing disorder by a well controlled thermal annealing schedule (Materials and Methods); it is inversely proportional to the electron localization length \cite{Shklovskii-Efros1984} $\xi$, which we will show controls the $g$-factor distribution width $\Delta g$.  We remark that in this regime (at low $T$) the detected magnetic moment appears `flat' in temperature (Fig.~ 1{B}).
	
	In the variable range hopping (VRH) regime, the fit of conductance  to Eq.~(1) for two states of disorder is illustrated in Fig.~2{C}. As seen in the figure, at low magnetic fields the characteristic non-analytic behavior is accurately followed; here the ratio of hopping time to spin relaxation time $\tau/\tau_s$ and the hopping field scale $H^{\star\star}$ were used as fitting parameters (Materials and Methods). The fits at different disorder levels controlled by the anneals at different temperatures $T_a$ are shown in {Fig.~S4}. The ratio $\tau/\tau_s$ strongly depends on the level of disorder  (Fig.~2{D}), with the hopping and spin relaxation rates, $1/\tau \propto H^{\star\star}$ and $1/\tau_s \propto H_s$, in close correspondence with the disorder dependence of $R_{xx}$ all the way through crystallization transition
	(see Fig. 5 below). The hopping time $\tau$ can be independently extracted from the E-S energy {(Fig.~S5A)}, and, as expected for the hopping conductivity $\tau$ increases exponentially on decreasing temperature ({Fig.~S5B}). This allows us to consistently obtain the evolution of spin-relaxation time $\tau_s$ (Fig.~2{E}) and $\Delta g$ (Fig.~3{A}) with decreasing disorder (increasing $T_a$), and hence that of the low-field spin-relaxation scale $H_s = \frac {\hbar}{\mu_B\tau_s\Delta g}$ {(Fig.~S6)}; $H_s$  marks a crossover from the concave-up field shape associated with $\tau_s$ to concave-down behavior at higher fields, see Fig.~2{C}.
	
	An intriguing question arises as to what controls the unexpectedly large (Tesla-range) field scale where negative magnetoresistance, the hallmark of dynamic spin memory, is found. For the material systems with small spin-orbit coupling where $g$-factor $\sim$2, the expected field range would be in the $10^{-4}-10^{-5}$ Tesla range. In theory \cite{Aleiner-Spivak2014}, this field scale is obtained from the competition of the magnetic energy of the spins, $g\mu_B H$, and either thermal energy or the exchange energy $J$ between neighboring spins -- it ought to be well below the competing effects.
	Here, however, with large effective $g$ value \cite{Wolos2016}, $g\mu_B H/k_B \sim 20$ {K} is comparable to the spin-memory range. This brings us back to disorder-induced spin correlations. Fig.~1{B} shows that under extreme disorder the onset of magnetic response is abrupt and at a remarkably high temperature $T_s \sim 200$ K. While the details of spin correlations in this Anderson-like-localized glassy state clearly deserve further experimental and theoretical studies, a rough estimate of $J \sim k_B T_s/z \approx 70$ {K}, using local coordination number \cite{LiaPCM2009} $z \sim 3$ expected in Sb$_2$Te$_3$, implies that here short range interactions between localized spins play the key role {(see SI, Section \textbf{A})}.
	
	\begin{figure*}[tb]
		\centering
		\includegraphics[width=\textwidth]{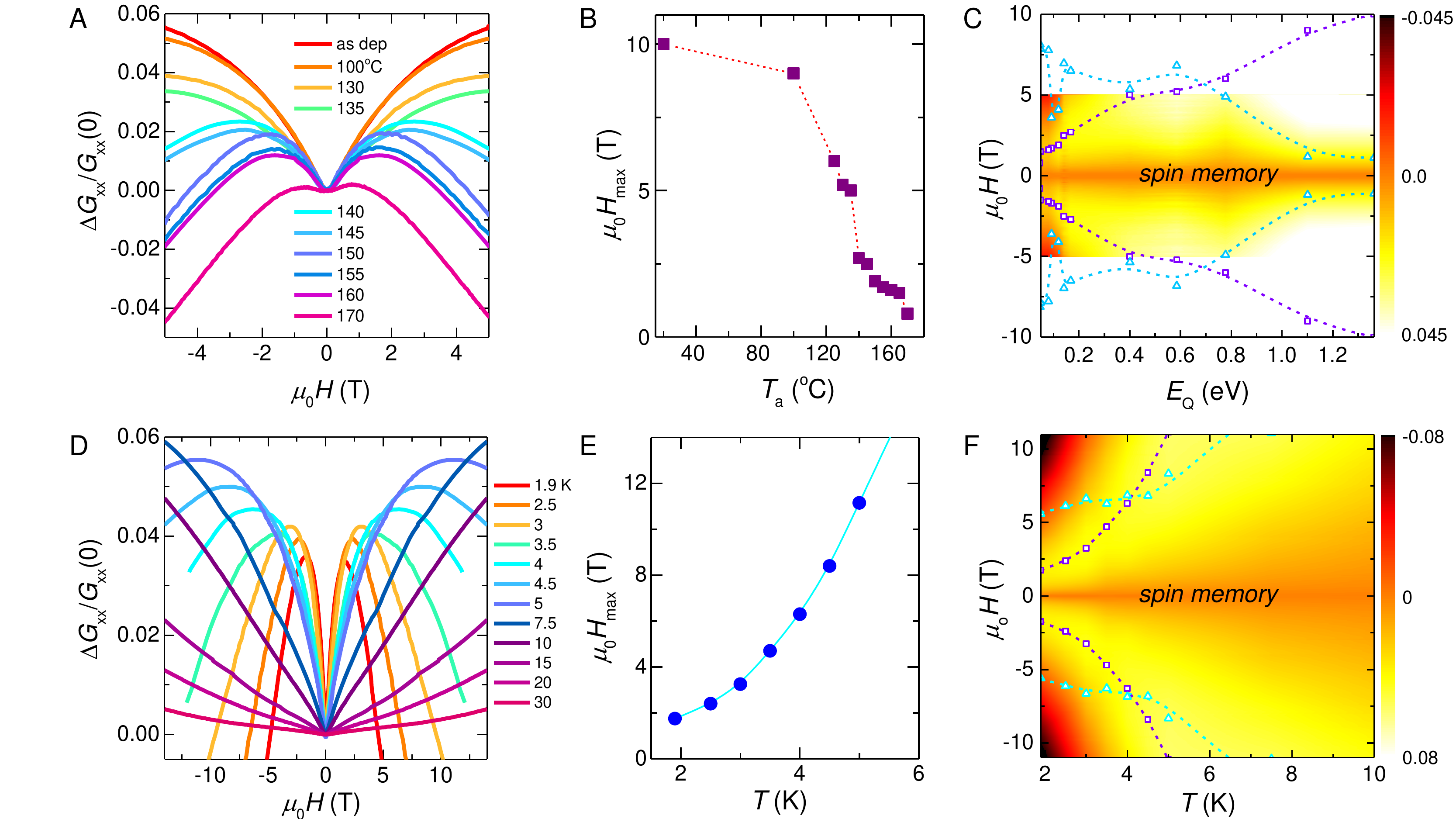}
		\caption{{Phase diagrams of the spin-memory state.}
			{(A)},  Normalized change in  $\Delta G_{xx}(H)/G_{xx}(0)$  at 1.9 K for different levels of disorder labeled by $T_{a}$. At low disorder the non-analytic positive spin-memory `cusp'  transforms into a parabolic form when the huge conductance rise near crystallization is nearly complete. {(B)},  The transformation occurs via reduction of the crossover field  $H_{max}$.  {(C)},  Phase diagram in the $H-E_Q$ space showing the huge disorder range where spin memory effect is present.
			{(D)},  $\Delta G_{xx}(H)/G_{xx}(0)$ shown for $T_{a} = 150^\circ\textrm{C}$ at different bath temperatures $T$.
			As $T$ increases the spin-memory `cusp' grows and widens until the system exits the VRH regime and enters the regime where transport
			eventually becomes diffusive.
			{(E)},   In the VRH range the crossover field $H_{max}$ monotonically increases with $T$.
			{(F)},  Phase diagram in the $H-T$ space controlled by spin memory. At low $T$ this range is bounded by $H_{max} \lesssim H^{\star\star}$. The color scales of the contour plots indicate the level of $\Delta G_{xx}(H)/G_{xx}(0)$. $H_{max}$ (hollow purple squares) and the hopping field scale $H^{\star\star}$ (hollow blue triangles) combine to define a limiting envelope for the spin-memory space.
		}
		\label{fig4}
	\end{figure*}
	
	\begin{figure}
		\includegraphics[width=1\linewidth]{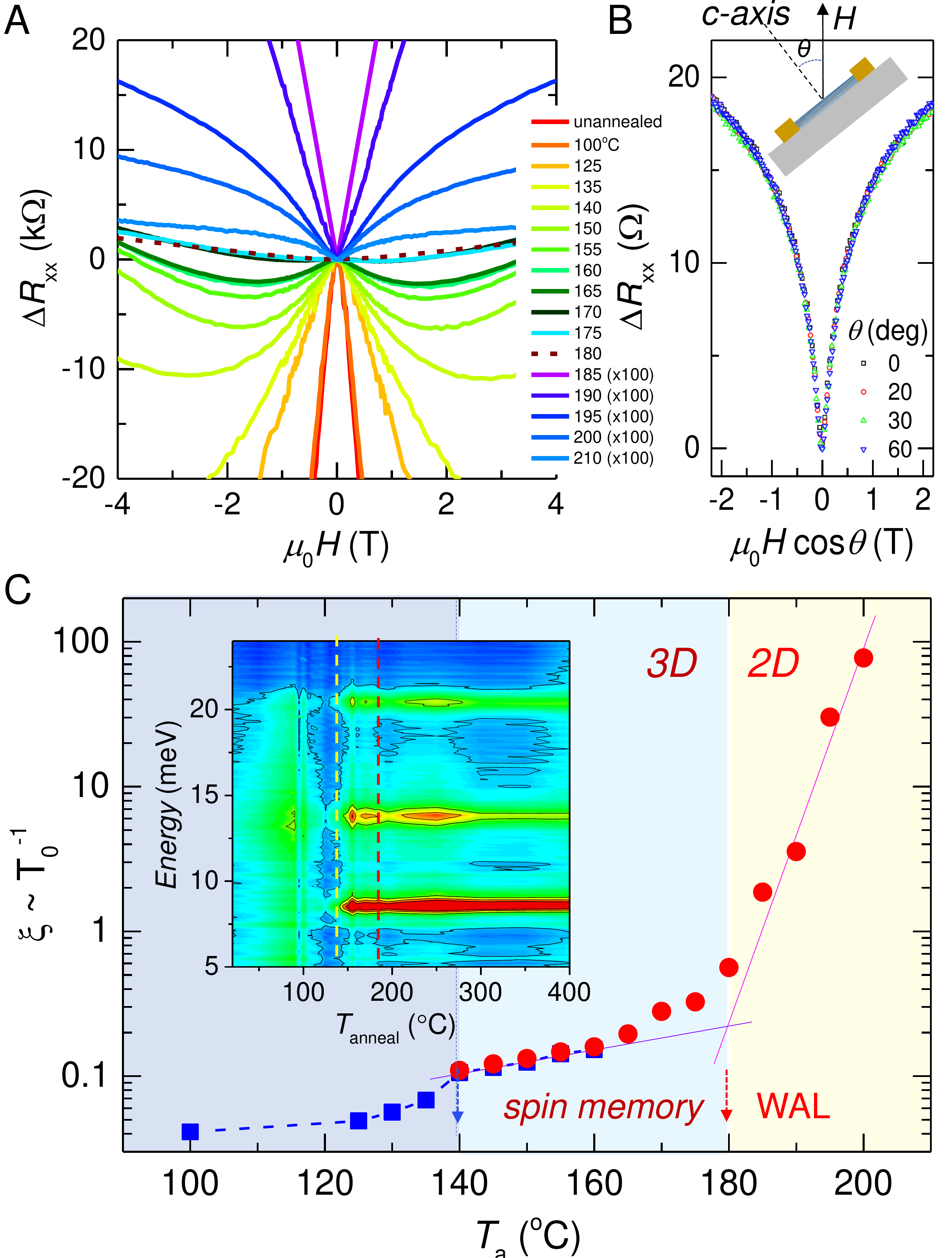}
		\caption{ {Transition from the spin-memory to weak antilocalization (WAL) state in disordered crystalline Sb$_2$Te$_3$ films.}
			{(A)},  Upon annealing, $\Delta R_{xx}$ undergoes transition from the spin memory neg-MR state to a positive magnetoresistance (pos-MR) WAL regime at $T_{a}\cong 180^\circ \textrm{C}$, when charge transport becomes diffusive.  {(B)},  The WAL pos-MR has a 2D character as evidenced by the scaling with the transverse component of magnetic field.  {(C)},  Localization length $\xi  \propto T_0^{-1}$ shows a distinct kink at crystallization and a smooth transition to a 2D WAL state. $T_0$ was obtained from fitting to the E-S VRH formula \cite{Gantmakher2005}. Inset: Crystallization at $T_{a}\cong 140^\circ \textrm{C}$  (yellow dash) is clearly seen as sharp lines in the Raman spectra \emph{vs.} $T_{a}$. Transition to WAL at $\simeq 180^\circ \textrm{C}$  is indicated by red dash.  }
		\label{fig5tc}
	\end{figure}
	Let us now consider $g$-factor fluctuations in a strongly disordered state. The $g$ distribution width naturally arising from our magnetoconductance data (Fig.~3{A}) within the model considered above is spectacularly wide at the highest level of disorder, $\Delta g \simeq 40$; indeed, it exceeds the effective $g$ value of $\sim 30$  obtained \textit{e.g.} directly in the same topological insulator family from the electron spin resonance (ESR) experiments \cite{Wolos2016}. Such large $g$ spread is uncommon but not unprecedented. Giant fluctuations of $g$-factors have been reported in \textit{e.g.} InAs nanowires where $\Delta g > |g_{eff}|$ and the effective factor $|g_{eff}|$ is also large \cite{Csonka2008}. Furthermore, in semiconducting quantum wells $|g_{eff}|$ has been known to increase roughly linearly \cite{Chen2007} with quantum confinement energy $E_Q$ as $|g_{eff}| \simeq g_0 + \beta E_Q$, where $\beta$ is a material-specific constant.  Here we propose that in the strongly localized Anderson-like state, quantum confinement is enforced by the wells constrained by the localization length $\xi$ (Fig.~3{B}). In this view, a simple particle-in-a-box approximation gives $E_Q \propto \xi^{-2}$ that indeed fully scales with $\Delta g$ (Fig.~3{A}). The expected linear $|g_{eff}|$ \textit{vs.} $E_Q$ would then 
	set  $|g_{eff}| \sim  \Delta g^{-\alpha}$ with $\alpha \sim 0.1$, with $|g_{eff}|$ approaching the ESR-determined value in the unconfined state, see Fig.~3{C}.
	
	Our experiments reveal that positive $\Delta G_{xx}(H)$ (neg-MR),
	evolves progressively with decreasing disorder;  it persists over a spectacularly large disorder range, all the way through the crystallization process and beyond (Fig.~4{A}). 
	The large limiting field range set by $H_{max}$ at strong disorder
	falls with increasing $T_a$ (Fig.~4{B}) to pinch off $\Delta G_{xx}(H)$ eventually to null.  A clear visual of the field-disorder phase space
	is shown in Fig.~4{C}, where the strength of disorder is represented by  $E_Q$. $H-E_Q$ diagram shows that
	spin-memory region is restricted by $H^{\star\star}$ to relatively low fields, but when the system is less localized the `envelope' of the spin-memory space switches to $H_{max}$.
	
	{The temperature range over which spin memory is evident is set by the VRH process (Fig.~2{B}), with
		$\Delta G_{xx}(H)$ well
		described by Eq. (1).
		Above $\sim 10$ {K} , outside the VRH region, 
		$\Delta G_{xx}(H)$  becomes nearly `flat' (Fig.~4{D}); there both $\tau_s$ ({Fig.~S7A}) and $|g_{eff}|$ ({see Fig.~S7B}) appear to saturate, and spin-memory phenomenon is not expected. The typical field scale associated with the E-S energy $T_0 \sim 10-30$ {K} is in the 1-2 T range, in close correspondence with $H^{\star\star}$ at low $T$.
		As before, $H_{max} \ncong H^{\star \star}$, and in the low-temperature localized state $H_{max}$ becomes the limiting crossover field (Fig.~4{E}).}
	Above $T_a   \simeq 180^\circ\textrm{C}$ ($E_Q \approx 60$ {meV}), in the strongly disordered \emph{crystalline} state spin memory phenomenon is not detectable.
	The exit from the spin-memory state at $\cong 180^\circ\textrm{C}$ is clearly evidenced by a transition from the negative MR (neg-MR) to a positive MR  (pos-MR)`cusp' characteristic of the WAL state (Fig. 5A). The WAL cusp  scales with the transverse component  of applied magnetic field $H_{\perp} = H\textrm{cos} \theta$ (Fig. 5B), consistent with the 2D (orbital) character  expected in a topological  insulator \cite{irrad-Lukas2016} under weak disorder. This 2D scaling should be contrasted with isotropic (3D) scaling of the neg-MR  peak in the spin-memory state.  Thus,  unlike the WL-WAL transition driven by magnetic impurities \cite{WL-WAL-Xue2012},  the transition from spin-memory onto a WAL state is also a 3D-2D dimensionality transition at which the electron system rapidly delocalizes. The change of the localization  length  $\xi $  with decreasing disorder is smooth at the transition to WAL (Fig. 5C) , with electron mean free path only limited by the film thickness and grain size in the crystalline state at high $T_a$ (Fig. S8). We remark again that WAL onsets at $ \simeq 180^\circ\textrm{C}$, way above the crystallization transition at  $  \simeq 140^\circ\textrm{C}$ (see inset in Fig. 5C),  pointing to  a disorder threshold for the topological state.
	
	To summarize, we directly detect spin response of the topological material Sb$_2$Te$_3$ under strong structural disorder, which we can control and tune by thermal annealing from amorphus to crystalline state. Under strong disorder, the system develops spin-correlations that drive the spin-memory phenomenon controlling the transport of charge. Both in magnetic field and in disorder strength, the parameter space where spin memory exists is unexpectedly broad --- it persists well into a disordered crystalline state where, eventually,  a 2D WAL quantum interference correction is recovered. While a simple phenomenology we used captures the key features of transport under strong disorder, this regime is undoubtedly complex and theory has yet to provide a full understanding of charge transport under such conditions, particularly when spin-orbit coupling is strong.  A new perspective revealed by our findings is that spin-memory effect sets a disorder threshold at which topological protection of the gapless surface states is reclaimed.
	Save for the edge currents, the control of spin-dependent transport arising from nonequilibrium correlations can, in principle, be achieved by using electrostatic gating 
	\cite{Sandhu2001}, and by modifying correlations with an addition of localized spins using currently practiced semiconductor doping techniques.
	
	\begin{acknowledgments}
		We wish to acknowledge Roland Winkler for his key insights regarding $g$-factors. We thank Igor Aleiner and Boris Spivak for their useful comments. We are grateful to Andy Kellock for the RBS and PIXE analysis of the films. This work was supported by the NSF grants DMR-1312483-MWN, DMR-1420634, and HRD-1547830 (L.K.-E.).
	\end{acknowledgments}
	
	\section{Appendix}
	\appendix{Film growth and structural characterization. Films of Sb$_2$Te$_3$ with thicknesses ranging from 20 to 100 nm were sputter-deposited at room temperature in Ar gas at 4 mTorr and a flow of 46 sccm from a nominally stoichiometric target using 15 W DC power on  Si$_3$N$_4$ (100 nm)/Si substrates.  The stoichiometry was confirmed by Rutherford Backscattering (RBS) and particle induced X-ray emission (PIXE). RBS data were collected at NEC 3UH Pelletron using a Si surface barrier detector with He$^+$ ions at 2.3 MeV. PIXE data was collected using a Si-Li detector with H$^+$ ions at 1 MeV. Elemental analysis was done at Evans Analytical Group. X-ray diffraction characterization was performed using Bruker D8 Discover system with the da Vinci configuration using a monochromated beam ($\lambda_{Cu} =1.5418 {\AA}$) and a scintillator detector with analyzer crystal (HR-XRD). The film morphology was characterized using the FEI Titan Themis 200 transmission electron microscope (TEM), 200 kV, with TEM resolution $0.9{\AA}$ and $4k \times 4k$ Ceta 16M CMOS camera. Disorder was characterized by Raman spectra  using 633 nm linearly polarized excitation in a backscattering configuration \cite{Raman-Secor-APL2014}, with power kept below 2 mW to avoid heating effects.

 Magnetic measurements were performed using custom-designed on-chip $\mu$-Hall sensors based on In$_{0.15}$Ga$_{0.85}$As heterostructures ({SI}). To measure magnetization, $\sim$50 micron in lateral size thin Sb$_2$Te$_3$ film samples were exfoliated from their substrates and placed directly on the SiO$_2$-passivated sensor using PDMS. At each temperature an empty twin sensor was used for background subtraction.
Transport measurements were performed in a 14 Tesla Quantum Design PPMS system in 1 mTorr of He gas on many film samples, each subjected to the same annealing protocol used to tune the level of disorder. Lithographically patterned structures combining both Hall bar and van der Pauw electrical contact configurations with Ti/Au metallurgy were used (Fig.~2A).  Measurements were performed on as-deposited films and on the same films after each 5 min annealing step in a box furnace in flowing nitrogen in the temperature range across crystallization at $T_a \sim 140^\circ\textrm{C}$. We used a numerical Monte Carlo technique to fit our transport data to Eq. (1).}
	
\bibliography{refsMain126c,booksetc5}

\end{document}